\documentclass[12pt]{article}
\usepackage{latexsym}
\usepackage{graphicx}
\usepackage{caption}
\usepackage{psfrag}
\usepackage{amsmath,amssymb,dsfont}
\newcommand{\be}{\begin{equation}}
\newcommand{\ee}{\end{equation}}
\newcommand{\bea}{\begin{eqnarray}}
\newcommand{\eea}{\end{eqnarray}}

\newcommand{\nn}{\nonumber}
\input epsf

\begin{document}

\begin{titlepage}

\begin{flushright}
%\today
\end{flushright}
\vspace*{1.5cm}
\begin{center}
{\Large \bf A rational approximation to $\langle VV-AA\rangle$ and its $\mathcal{O}(p^6)$
low-energy constant}\\[3.0cm]

{\bf Pere Masjuan} and {\bf Santiago Peris}\\[1cm]

Grup de F{\'\i}sica Te{\`o}rica and IFAE\\ Universitat Aut{\`o}noma de Barcelona, 08193 Barcelona, Spain.\\[0.5cm]

\end{center}

\vspace*{1.0cm}

\begin{abstract}

Using a sequence of rational approximants and the large-$N_c$ limit of QCD, we estimate the value of the
low-energy constant $C_{87}$ which appears in the Lagrangian of Chiral Perturbation Theory at
$\mathcal{O}(p^6)$.

\end{abstract}

\end{titlepage}

The Chiral Lagrangian\cite{Weinberg,GL} organizes the physics of the strong interactions at low energy as an
expansion in powers of  momentum and masses of the lightest pseudoscalar fields, which are the only ones
explicitly present in this Lagrangian. Since all the heavier states of QCD are integrated out, their physics
is encoded in a set of low-energy constants (LECs). These LECs are indispensable to make definite
predictions in Chiral Perturbation Theory. There is already a relatively good knowledge of the value of most
of the LECs which appear at $\mathcal{O}(p^4)$ in the chiral expansion\cite{GL, Bernard}. At
$\mathcal{O}(p^6)$, however, most of the $\mathcal{O}(100)$ LECs are completely unknown. This note is
concerned with the estimate of one of them, the LEC of $\mathcal{O}(p^6)$ $C_{87}$\cite{p6} appearing in the
$<VV-AA>$ two-point correlator in the chiral limit.

The general strategy will consist in constructing a rational approximant to the relevant Green's function,
i.e. $<VV-AA>$, from the coefficients of the chiral expansion and any other known properties of the full
function. Once the rational approximant is known, upon reexpansion around $Q^2=0$, higher order unknown
coefficients of the chiral expansion may be predicted. If the rational approximant is a better description
of the original function than the partial sums of the chiral expansion, one may expect this prediction to be
reliable. For a brief review, where further references to the literature may be found, we refer to
Weniger\cite{Weniger}.

 Let us, therefore, consider the two-point  functions of vector and axial-vector
currents in the chiral limit of QCD\be \label{correlator} \Pi^{V,A}_{\mu\nu}(q)=\ i\,\int
d^4x\,e^{iqx}\langle J^{V,A}_{\mu}(x) J^{\dag\ V,A}_{\nu}(0)\rangle = \left(q_{\mu} q_{\nu} - g_{\mu\nu} q^2
\right)\Pi_{V,A}(q^2) \ , \ee with $J_{V}^\mu(x) = {\overline d}(x)\gamma^\mu u(x)$ and $J_A^\mu(x) =
{\overline d}(x)\gamma^\mu \gamma^5 u(x)$. As it is known, the difference $\Pi_{V-A}(q^2)$ satisfies the
unsubtracted dispersion relation given by\footnote{The upper cutoff which is needed to render the dispersive
integrals mathematically well defined can be sent to infinity provided it respects chiral symmetry
\cite{GP02}.} \be \label{dispersion} \Pi_{V-A}(q^2)\equiv \frac{1}{2}(\Pi_{V}(q^2)-\Pi_{A}(q^2))=
\lim_{\Lambda\rightarrow\infty}\int_0^{\Lambda^2} \frac{dt}{t-q^2-i\epsilon}\ \frac{1}{\pi}\ {\rm
Im}\,\Pi_{V-A}(t)\ . \ee

Since all LECs are defined in the chiral limit, the restriction of the function $\Pi_{V-A}(q^2)$ to this
limit entails no loss of generality. Even then, the analytic structure of $\Pi_{V-A}(q^2)$ is very
complicated, with a multiparticle cut starting at $t\geq 0$. A further simplification occurs in the
large-$N_c$ limit of QCD\cite{thooft} in which the previous cut becomes suppressed and only single particle
intermediate states are allowed. The function $\Pi_{V-A}(q^2)$ contains then an infinite set of isolated
poles \cite{Witten}, and becomes meromorphic. In Ref. \cite{us} it was emphasized that any truncation of
this meromorphic function to a finite set of poles may be interpreted as a rational approximation to the
original function. There are a number of reasons why the large-$N_c$ limit of QCD is a sensible limit to
take, in particular for Green's functions built out of the vector and axial-vector currents
\cite{Jaffe,Witten}.

Therefore, in the large-$N_c$ limit, the function $q^2 \Pi_{V-A}(q^2)$ has the following
representation\footnote{Multiplication by $q^2$ kills the pion pole at the origin.}
\begin{equation}\label{largeN}
    q^2 \Pi_{V-A}(q^2)=f_0^2+ q^2 \sum_{R=0}^{\infty}\frac{c_R}{-q^2+M^2_R}
\end{equation}
where $R$ labels resonance states and, assuming the existence of a mass gap, all masses are nonzero with the
rho meson mass being the smallest one in the sum. In this way,  defining $Q^2=-q^2$, the expression
(\ref{largeN}) is analytic at $Q^2=0$, and allows the Taylor expansion\cite{ABT}
\begin{equation}\label{largeNtaylor}
    - Q^2 \Pi_{V-A}(-Q^2)\approx f^2_{0}-4 \ L_{10}\ Q^2 -8\  C_{87}\ Q^4 + \ldots \quad , \quad (Q^2\rightarrow
    0)
\end{equation}
where $f^2_{0}$ is the pion decay constant (in the chiral limit), and  $L_{10}, C_{87} $ are the LECs at
$\mathcal{O}(p^4,p^6)$ (respectively) in the corresponding Chiral Lagrangian\cite{GL,p6}. On the other hand,
the expansion of $\Pi_{V-A} $ at infinity is
\begin{equation}\label{largeNinf}
    -Q^2 \Pi_{V-A}(-Q^2)\approx 4 \pi \alpha_s \frac{\langle\overline{\psi}\psi\rangle^2}{Q^4}\bigg(1+ \mathcal{O}
    \left( \alpha_s \log Q^2 \right) \bigg)+ \ldots \quad , \quad (Q^2\rightarrow
    \infty)
\end{equation}
where $\alpha_s$ is the QCD coupling constant and $\langle\overline{\psi}\psi\rangle$ is the quark
condensate. Unlike the expansion around the origin, the existence of nonvanishing anomalous dimensions, even
in the large-$N_c$ limit, gives rise to the $\log Q^2$ terms and, unlike (\ref{largeNtaylor}), renders the
expansion around infinity in (\ref{largeNinf}) \emph{not} analytic.

The expansion (\ref{largeNtaylor}) will be our starting point in this work. Although the LEC $L_{10}$ is
pretty well known\cite{GL,BijnensLis}, this is not so for $C_{87}$. It is therefore important to obtain a
new determination of this LEC with its associated error.

Given the meromorphic function $Q^2 \Pi_{V-A}(-Q^2)$ in the $Q^2$ complex plane with an analytic expansion
around the origin, as in (\ref{largeNtaylor}), it is possible to construct a Pade Approximant (PA),
$P^{M}_{N}(Q^2)$, as the ratio of two polynomials of degree $M$ and $N$ (respectively) such that its
expansion in powers of $Q^2$ matches that of the original function up to, and including, the term of
$\mathcal{O}(Q^{2(M+N)})$. Since the function falls off at large $Q^2$ as $Q^{-4}$ up to logarithms (see
(\ref{largeNinf})), we choose  $N=M+2$ in order to optimize the matching of the rational approximant at
large $Q^2$ to this behavior\footnote{Due to the presence of logarithms in (\ref{largeNinf}), however, this
matching cannot be perfect.}. We emphasize, however, that this choice does not affect the properties of
convergence of Pade Approximants, as described next.

As $M\rightarrow \infty$, there is a theorem\cite{Pommerenke} that ensures convergence of the sequence of
PAs to the original meromorphic function, in any compact set in the complex $Q^2$ plane except at a finite
number of poles. Of course, where there is convergence, the PA may be considered an approximate resummation
of the Taylor series around the origin. On the other hand, the set of points where there is no convergence
certainly includes the position of the poles since not even the original function is defined there, but
there may appear other artificial poles which have no counterpart in the original function. One would
naively think that the presence of these artificial poles would cause a major distortion and completely
spoil the rational approximation. However, one can actually show\cite{Pommerenke} that as the order of the
Pade increases, i.e. as $M$ grows, these artificial poles either move to infinity in the complex plane and
decouple or they get "almost-canceled" by the appearance of nearby zeros. Although, in general, this
cancelation is not complete, it is efficient enough to make the region of distortion of the artificial pole
only of zero measure. This is why and how the Pade Approximation works. For an explicit example where all
these properties come to play in the context of a Regge-inspired model, we refer to \cite{mp}.

Theorem \cite{Pommerenke} is important because it teaches us useful information about the qualitative
behavior of how Pades approximate meromorphic functions. Regretfully, when asking more quantitative
questions such as the rate of convergence, which is the first step towards an estimate of the error, such a
theorem is only of limited practical importance. In practice, one can take a more useful approach towards an
estimate of the error by studying the behavior of a set of successive rational approximants, as we will now
explain.

In order to be able to construct a sequence of rational approximants it is of course crucial to have enough
number of inputs. Since PAs are constructed from the coefficients of the Taylor expansion
(\ref{largeNtaylor}) one immediately faces an obvious difficulty. Since what one wishes is an estimate of
$C_{87}$, only the two coefficients $f_0$ and $L_{10}$ may be used. With these two coefficients as input,
the only PA vanishing at large $Q^2$ is $P^{0}_{1}$, but it falls off as $Q^{-2}$ which is too slow as
compared to (\ref{largeNinf}). Consequently, it is  necessary to consider more general rational approximants
than the standard PAs.

In Ref. \cite{mp} we saw that the $\pi^+-\pi^0$ mass difference in the chiral limit, which is given by
\begin{equation}\label{pimass}
    \delta M^{2}_{\pi}\equiv \ M^2_{\pi^{+}}- M^2_{\pi^{0}}= - \frac{3}{4\pi}\ \frac{\alpha}{f_0^2}\ \int_{0}^{\infty}
    dQ^2\ Q^2\ \Pi_{V-A}(-Q^2)\ ,
\end{equation}
could be used as a further constraint in the construction of the PAs with very good numerical results.
Together with $f_0$ and $L_{10}$ one now has three inputs to construct the $P^{0}_{2}$, which does match the
power fall-off at large $Q^2$ in (\ref{largeNinf}). By simple re-expansion around $Q^2=0$ it is then
possible to predict an estimate for the term of $\mathcal{O}(Q^4)$ in (\ref{largeNtaylor}). This prediction
was checked against the exact value in the model in \cite{mp} with very good results, and this encouraged us
to do the same also for QCD. In the QCD case, using the values\footnote{Since $m_s$ decouples from $F_{\pi}$
in the large-$N_c$ limit, the value of $f_0$ is estimated in Eq. (\ref{realnature}) by extracting the chiral
corrections from $F_{\pi}$ using $SU(2)\times SU(2)$ chiral perturbation theory, but doubling the error as
compared to Ref. \cite{Colangelo}. }
\begin{eqnarray}\label{realnature}
f_0&=& 0.086\pm  0.001\ \mathrm{GeV} \quad ,\qquad  \mathrm{( Ref.\ \cite{Colangelo})} \nn\\
   \delta M_{\pi}&=& 4.5936\pm 0.0005\ \mathrm{MeV}\ ,\qquad \mathrm{(Ref.\ \cite{PDG})}  \\
\!\!\!\!\!\!\!\!\!\!\!\!\!\!L_{10}(0.5\ \mathrm{GeV}) \!\leq \!L_{10}\!\leq \!L_{10}(1.1\
\mathrm{GeV})&\!\!\!\! \Longrightarrow \!\!\!\!& L_{10}=\left( -5.13\pm 0.6\right) \times 10^{-3} \ ,\
\mathrm{(Ref.\ \cite{Stern})}\ , \nn
\end{eqnarray}
we then obtained \cite{mp} an estimate of the $\mathcal{O}(Q^4)$ term in the expansion (\ref{largeNtaylor})
which translates into the value $C_{87}=  (5.4 \pm 1.6 )\times 10^{-3}\ \mathrm{GeV^{-2}}$. In the present
work, we would like to reassess this value with a more complete analysis.

There are different kinds of rational approximants closely related to the usual PAs which, although perhaps
not so popular, are also very useful. Among those, we would like to stress the so-called Pade-Type
Approximants (PTA)\cite{PGV}, $\mathbb{T}^M_N(Q^2)$, which are very useful when one has some knowledge of
the spectrum of resonance masses in the original Green's function. They are defined as the ratio of two
polynomials $Q_M(Q^2)$ and $T_N(Q^2)$ (of degrees $M$ and $N$, respectively):
\begin{equation}\label{five}
    \mathbb{T}^M_N(Q^2)=\frac{Q_M(Q^2)}{T_{N}(Q^2)}\ ,
\end{equation}
where the polynomial in the denominator has its $N$ zeros preassigned precisely at the positions of the
first $N$ resonance masses in the original Green's function (\ref{largeN}), i.e.
\begin{equation}\label{TN}
    T_{N}(Q^2)= (Q^2+M_1^2)  (Q^2+M_2^2)...(Q^2+M_N^2)\ ,
\end{equation}
and the polynomial $Q_M(Q^2)$ is defined so that the expansion of the PTA around $Q^2=0$ agrees with that of
the original function up to terms of order $M+1$, i.e.
\begin{equation}\label{six}
    \mathbb{T}^M_N(Q^2)\approx f_0 + f_1\ Q^2 + f_2\ Q^4
    +...+ f_{M}\ Q^{2M}+ \mathcal{O}(Q^{2(M+1)})\ .
\end{equation}
Choosing $N=M+2$ one  optimizes the matching of $\mathbb{T}^M_{M+2}(Q^2)$ to the expansion (\ref{largeNinf})
at $Q^2\rightarrow \infty$ and, as in the case of PAs, this is a choice we will make.

Both PTAs and PAs where studied in Ref. \cite{mp} and  the lessons which can be drawn from that model are
the following. The model confirms that one may estimate the unknown LECs with these rational approximants
where, in the case of PTAs, the physical masses were chosen in increasing order, i.e. $M_1< M_2<M_3 ...$ For
instance, with the PTA $\mathbb{T}^M_{M+2}(Q^2)$ we could see that one has a good prediction for the term of
$\mathcal{O}(Q^{2(M+1)})$ in the low-$Q^2$ expansion, which is the first one not used as input, with a
precision which improves as the order of the approximant, $M$, increases. Furthermore, the accuracy obtained
for the unknown coefficients of the Taylor expansion is very hierarchical: the accuracy obtained for the
term $\mathcal{O}(Q^{2(M+1)})$ is better than that for the term of $\mathcal{O}(Q^{2(M+2)})$, and that
better than for the term $\mathcal{O}(Q^{2(M+3)})$, with a quick deterioration for higher-order terms. The
case of PAs follows the same pattern. As to the description of the spectrum, we found that PAs also
reproduced the values for the residues and masses in a hierarchical way: while the first masses and residues
are well reproduced, the prediction quickly worsens so that the last pole and residue of the PA has no
resemblance whatsoever with its physical counterpart. The same is true for the residues of a PTA (since the
masses are fixed to be the physical ones by construction).

Based on the above, one can envisage the following strategy for getting a sequence of estimates for the
$\mathcal{O}(p^6)$ LEC $C_{87}$. Assuming that the vector and axial-vector meson masses stay approximately
the same in the large-$N_c$ and chiral limits, one can use their values extracted from the PDG
book\cite{PDG} to construct several PTAs. We think that this assumption is reasonable for both limits.
First, for the chiral limit, this is because the up and down quark masses are very small\cite{pelaez}.
Second, for the large-$N_c$ limit, there is a non negligible amount of phenomenological evidence in favor of
the rho meson being a $q\overline{q}$ state\cite{pelaez2,Jaffe}. Besides, the success in the spectroscopy of
the quenched lattice results for the lightest vector mesons is also suggestive that $1/N_c$ corrections may
not be very large\cite{Sharpe}\footnote{Be that as it may, whether the assumption is correct or not will
ultimately be judged by the final results obtained.}. Therefore, we will use for the masses
\begin{eqnarray}\label{masses}
  m_{\rho}= 0.7759 \pm 0.0005\ ,
  m_{\rho'}=1.459 \pm 0.011 &\!\!\!\!\!\!\!\!\!,&\!\!\!\!\!\!\!\!\!\!\!\  m_{\rho''}=1.720 \pm 0.020,
  m_{\rho'''}= 1.880 \pm 0.030 \nn \\
   m_{a_1}=1.230 \pm 0.040&,&\  m_{a'_1}= 1.647 \pm 0.022 ,
\end{eqnarray}
where all the numbers have been expressed in GeV.

For instance, with only $f_0^2$ and the masses of the $\rho$ and $a_1$, one can construct the PTA
$\mathbb{T}^0_{2}(Q^2)$ and predict the value for $L_{10}=(-4.32 \pm 0.02)\times 10^{-3}$, which is not bad
when compared, e.g., with (\ref{realnature}). The next term in the expansion gives the following value for
$C_{87}= (4.00 \pm 0.09)\times 10^{-3}\ \mathrm{GeV^{-2}}$ which is similar to that obtained in \cite{mp}
with the Pade $P^{0}_{2}$. However, since this value for $C_{87}$ comes from the second unknown term in the
expansion of $\mathbb{T}^0_{2}(Q^2)$ rather than the first, it is quoted here only for illustrative purposes
and will not be included in our final estimate, in agreement with our previous discussion. Adding $L_{10}$
and the $\rho'$ mass to the previous set of inputs one can then construct $\mathbb{T}^1_{3}(Q^2)$, which
produces $C_{87}= (5.13 \pm 0.26 )\times 10^{-3}\ \mathrm{GeV^{-2}} $. The PTA $\mathbb{T}^{2}_{4}$ can be
constructed if one also uses the pion mass difference (\ref{pimass}) and $m_{a'_1}$, yielding in this case
$C_{87}= (5.24 \pm 0.33)\times 10^{-3}\ \mathrm{GeV^{-2}} $. We find the stability of these predictions
quite reassuring.

A comment on the quoted error estimates is in order. These quoted errors are the result of the propagation
of errors from the input via the montecarlo method\cite{Eadie}. As such, they do not reflect the intrinsic
systematic error due to the approximation itself which will be estimated, at the end,  as the spread of
values obtained with the sequence of different approximants. On the other hand, the propagation of the error
from the input via the montecarlo method consists in the following. Taking each input in (\ref{realnature})
and (\ref{masses}), we have constructed a sample of data with a gaussian probability distribution yielding
as the average and standard deviation precisely the corresponding input value and its quoted error,
respectively. For each member of this sample, the rational approximant is then constructed and, upon
reexpansion, the LEC is obtained. The distribution of the different values for $C_{87}$ so obtained happens
to be also gaussian to a very good approximation. Therefore it will have an average value $X$ and a standard
deviation $Y$ which are then used to quote the result for $C_{87}$ as $X\pm Y$.

To be able to construct further rational approximants one needs an extra assumption. Although, as we have
emphasized above, the residues of the heaviest poles in a rational approximant do not come out anywhere
close to the corresponding physical decay constants, this is not true for the lightest ones. In particular,
in Ref. \cite{mp}, it was seen that the value of the residue for the  first pole in a PTA could reproduce
the exact value in the model with very good precision \emph{if} the order of the PTA was high enough and,
more importantly, it was improving as the order of the PTA grows. Consequently, if we are willing to use the
decay constant $F_{\rho}$, and perhaps also the $F_{a_1}$, one can go for the construction of higher PTAs.
These two residues can be gotten from the decays $\rho\rightarrow e^+ e^-$ and $a_1\rightarrow \pi \gamma$,
respectively, and their values are\cite{decays}
\begin{equation}\label{decay}
    F_{\rho}=0.156 \pm 0.001 \quad  F_{a_1}=0.123 \pm 0.024
\end{equation}
in GeV units.

\begin{table}
\centering
\begin{tabular}{|c|||c|}%{lr@{\hspace{0.2em}}c@{\hspace{0.2em}}lc}
\hline
$\mathds{T}^n_m$ & inputs   \\
\hline \hline \hline
$\mathds{T}^0_2$ &     $f_{0}$ ; $m_{\rho}$, $m_a$  \\
\hline
$\mathds{T}^1_3$ &     $f_{0}$, $L_{10}$ ;  $m_{\rho}$, $m_a$, $m_{\rho^{'}}$   \\
\hline \hline
$\mathds{T}^{2\, (a)}_4$ &    $f_{0}$, $L_{10}$, $\delta M_{\pi}$ ;  $m_{\rho}$, $m_a$, $m_{\rho^{'}}$, $m_{a^{'}}$   \\
\hline
$\mathds{T}^{2\, (b)}_4$ &     $f_{0}$, $L_{10}$, $F_{\rho}$ ;  $m_{\rho}$, $m_a$, $m_{\rho^{'}}$, $m_{a^{'}}$   \\
\hline \hline
$\mathds{T}^{3\, (a)}_5$ &     $f_{0}$, $L_{10}$, $F_{\rho}$, $\delta M_{\pi}$ ; $m_{\rho}$, $m_a$, $m_{\rho^{'}}$, $m_{a^{'}}$, $m_{\rho^{''}}$   \\
\hline
$\mathds{T}^{3\, (b)}_5$ &    $f_{0}$, $L_{10}$, $F_{\rho}$, $F_a$ ;  $m_{\rho}$, $m_a$, $m_{\rho^{'}}$, $m_{a^{'}}$, $m_{\rho^{''}}$  \\
\hline \hline $\mathds{T}^{4}_6$ &     $f_{0}$, $L_{10}$, $F_{\rho}$, $F_a$, $\delta M_{\pi}$ ; $m_{\rho}$,
$m_a$, $m_{\rho^{'}}$,
$m_{a^{'}}$, $m_{\rho^{''}}$, $m_{\rho^{'''}}$   \\
\hline
\end{tabular}
\caption{Set of inputs used for the construction of the different Pade Type Approximants in the
text.}\label{Tab:PadeTypes}
\end{table}

For instance, using $f_0^2, L_{10}, \delta M^2_{\pi}$ and $F_{\rho}$, as well as the five masses $m_{\rho},
m_{a_1}, m_{\rho'}, m_{a'_1}$ and $m_{\rho''}$, one can construct the PTA $\mathbb{T}^{3}_{5}$. Upon
expanding this approximant, one obtains the value $C_{87}=(5.78\pm 0.21)\times 10^{-3}\ \mathrm{GeV^{-2}} $.
Alternatively, one can also use $f_0^2, L_{10}, F_{\rho}$ and only the first four masses to construct a
$\mathbb{T}^{2}_{4}$ approximant, which is different from the other $\mathbb{T}^{2}_{4}$ considered above.
The value obtained for $C_{87}$, i.e.  $C_{87}=(6.00\pm 0.15)\times 10^{-3}\ \mathrm{GeV^{-2}} $, is
nevertheless very similar, which again brings confidence on the prediction.

In this way we have constructed a variety of rational approximants which we have listed on Table 1, in
increasing order of the degree in the denominator, together with the set of inputs used. We have gone all
the way up until the $\mathds{T}^{4}_6$, with the six masses listed on (\ref{masses}).

Figure 1 shows the prediction for the LEC $C_{87}$ from the corresponding rational approximant shown on the
abscissa, upon expansion around $Q^2=0$. We also included our previous result obtained in Ref. \cite{mp}
with the PA $P^0_2$, but with the present montecarlo method for the treatment of errors. As one can see, the
stability of the result is quite striking. After averaging over all these points, we obtain as our final
result in the large-$N_c$ limit,
\begin{equation}\label{result}
    C_{87}= (5.7 \pm 0.5)\times 10^{-3}\ \mathrm{GeV^{-2}}.
\end{equation}
The error in (\ref{result}) is mainly dominated by the error on the input for $L_{10}$ in Eq.
(\ref{realnature}) and is rather insensitive to the errors on the other inputs. For instance, one could
increase the error on $f_0$ to 5 MeV in Eq. (\ref{realnature}), or the error on $m_{\rho}$ to 50 MeV in
(\ref{masses}), or the error on $\delta M_{\pi}$ to 0.5 MeV  in (\ref{realnature}), without falling out of
the error band given in (\ref{result}).

For comparison, we also show in Fig. 1 the result of several previous estimates for this LEC. Reference
\cite{ABT} (shown as `A') uses the residues in Eq. (\ref{decay}) and the $\rho$ and $a_1$ physical masses to
construct, in effect, what we could call the PTA $\mathbb{T}^{2}_{2}$ to $Q^2 \Pi_{V-A}$. The difference
between this result and ours stems from the fact that this rational approximant falls off like a constant at
large $Q^2$, unlike Eq. (\ref{largeNinf}). Also, as we have already emphasized, the use of the physical
decay constant $F_{a_{1}}$ (\ref{decay}) in a rational approximant which has the $a_1$ as the heaviest pole
is a potential source of error.

Reference \cite{Knecht} also obtains an estimate for $C_{87}$ (shown as `B') based on the construction of a
rational approximant which effectively coincides with the PTA $\mathbb{T}^{0}_{2}$ but using the physical
value of $F_{\pi}=92.4$ MeV \cite{PDG} instead of the value of $f_0$ in Eq. (\ref{realnature}). Had they
used $f_0$, the result would have been lower, and would have agreed with the value we mentioned in the
paragraph right after Eq. (\ref{masses}). Therefore, our comments on the Pade Type $\mathbb{T}^{0}_{2}$
found in that paragraph also apply to this determination in \cite{Knecht}.

Finally, one can get still another estimate for $C_{87}$ from the PTA $\mathbb{T}^{0}_{2}$ in \cite{Knecht}
by assuming that the $a_1$ mass in the large-$N_c$ limit is not approximated by the physical value in Eq.
(\ref{masses}), but by a value which comes from the radiative pion decay saturated with the $\rho$ and the
$a_1$. This value turns out to be $m_{a_{1}}\sim 998$ MeV \cite{Portoles}. This lower number for the $a_1$
mass is the reason for a higher value for $C_{87}$ than that obtained in \cite{Knecht}, and is shown as `C'
in Fig. 1. However, there is no compelling reason to associate this different mass of the $a_1$ with the
large-$N_c$ limit. In fact, our results show how similar values for $C_{87}$ can be obtained with the
physical masses of the mesons used for the poles. Morevover, one of the advantages of our method is that one
can get a rough idea about the systematic error involved  by looking at the dispersion of the values
obtained.

\begin{figure}
  \includegraphics[width=5in]{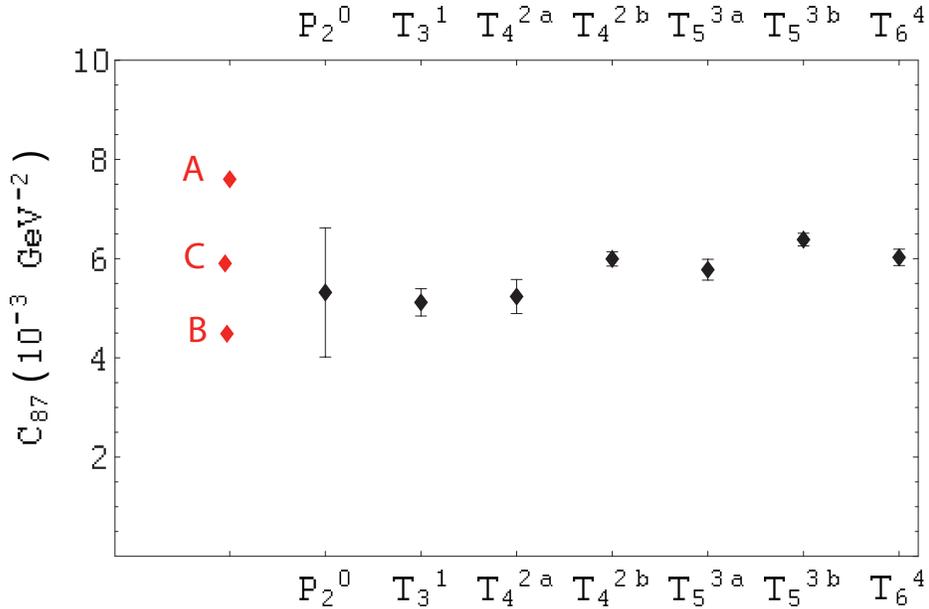}\\
  \caption{Prediction for $C_{87}$ in the large-$N_c$ limit from  the PA $P^0_2$ in Ref. \cite{mp}, and the different PTAs
  discussed in the text and appearing in Table 1.
   For comparison we also show the estimate from Refs. \cite{ABT,Knecht,Portoles},
   which we label `A', `B' and `C' (resp.).}\label{C4graph}
\end{figure}

Of course, in the large-$N_c$ limit $C_{87}$ does not run with scale whereas in the world at $N_c=3$ it
does. This is an additional source of systematic error in the result (\ref{result}). However,
phenomenological evidence as well as theoretical prejudice\cite{swiss} suggests that a reasonable guess for
this systematic error may be obtained by varying the scale in $C_{87}(\mu)$ between the range $0.5\
\mathrm{GeV}\lesssim \mu \lesssim 1\ \mathrm{GeV}$ (compare with $L_{10}$ in (\ref{realnature})). Using the
running obtained in Ref. \cite{run}, this error turns out to be $\sim 30$ per cent, right in the ballpark
expected for a typical $1/N_c$ effect. This systematic error should be added to our large-$N_c$ result in
Eq. (\ref{result}) in order to obtain an estimate for $C_{87}(\mu \sim 0.7)$ in the real world. In this
case, all the different results in Fig. 1 can be encompassed by this error.

We would like to finish by recalling that PAs and PTAs are, in a way, two extreme versions of a rational
approximant. While in the latter all poles are fixed at the physical masses, in the former the poles are
left free, and they are obtained by demanding that the expansion around $Q^2=0$ reproduces that of the
original function to the highest possible order. Besides these two rational approximants, there are also the
so-called  Partial Pade Approximants\cite{mp,PGV} which, from a certain point of view, lie half way between
PAs and PTAs. These Partial Pades are rational functions whose polynomial in the denominator has only some
of the poles preassigned but the others are left free, to be determined by the usual matching conditions at
$Q^2=0$. Therefore, there is no reason why, in general, the poles of a Partial Pade should come out to be
purely real\footnote{Although, when complex, they always come in complex conjugate pairs. This just means
that, in general, the poles of a rational approximant are not necessarily physical.}, unlike those of a PTA,
which are of course real by construction. We have constructed seven of these Partial Pades, with a
polynomial in the denominator up to fifth order in $Q^2$. In some of the cases the poles were actually
complex, as it was also the case of the PA $P^0_2$\cite{mp}. However, the results obtained for $C_{87}$ are
almost identical to those in Fig. 1, although with errors which are somewhat larger. This feature reinforces
the stability of the result shown in Fig. 1, and gives us reassurance about the reliability of our result.
Finally, we would like to mention that predicting $\mathcal{O}(p^8)$ LECs may also be another
straightforward application of this method.

\vspace{1cm}

\textbf{Acknowledgements}

We are grateful to M. Martinez for his instructive comments on the treatment of errors via the montecarlo
method and to J.J. Sanz-Cillero for discussions. One of us (SP) is grateful to P. Gonzalez Vera, G. Lopez
Lagomasino and J. Weniger for their explanations on the mathematical properties of Pade Approximants, and to
C. Brezinski for a useful correspondence on the same subject. This work has been supported by
CICYT-FEDER-FPA2005-02211, SGR2005-00916, the Spanish Consolider-Ingenio 2010 Program CPAN (CSD2007-00042)
and by the EU Contract No. MRTN-CT-2006-035482, ``FLAVIAnet''.


\begin{thebibliography}{99}

\bibitem{Weinberg}
 S.~Weinberg,
  %``Phenomenological Lagrangians,''
  Physica A {\bf 96} (1979) 327.
  %%CITATION = PHYSA,A96,327;%%


\bibitem{GL}
 J.~Gasser and H.~Leutwyler,
  %``Chiral Perturbation Theory: Expansions In The Mass Of The Strange Quark,''
  Nucl.\ Phys.\  B {\bf 250} (1985) 465.
  %%CITATION = NUPHA,B250,465;%%

\bibitem{Bernard}
C.~Aubin {\it et al.}  [MILC Collaboration],
  %``Light pseudoscalar decay constants, quark masses, and low energy  constants
  %from three-flavor lattice QCD,''
  Phys.\ Rev.\  D {\bf 70} (2004) 114501
  [arXiv:hep-lat/0407028].
  %%CITATION = PHRVA,D70,114501;%%

\bibitem{Weniger}
E.J. Weniger, J. Math. Phys. {\bf 45} (2004) 1209 [arXiv:math-ph/0306063], section 6.

\bibitem{GP02}
M.~Golterman and S.~Peris,
  %``On the use of the operator product expansion to constrain the hadron
  %spectrum,''
  Phys.\ Rev.\  D {\bf 67} (2003) 096001
  [arXiv:hep-ph/0207060].
  %%CITATION = PHRVA,D67,096001;%%


\bibitem{thooft}
G.~'t Hooft,
  %``A PLANAR DIAGRAM THEORY FOR STRONG INTERACTIONS,''
  Nucl.\ Phys.\  B {\bf 72} (1974) 461.
  %%CITATION = NUPHA,B72,461;%%

\bibitem{us}
S.~Peris,
  %``Electroweak matrix elements at large N(c): Matching quarks to mesons,''
  arXiv:hep-ph/0204181;
  %%CITATION = HEP-PH 0204181;%%
E.~de Rafael,
  %``Analytic approaches to kaon physics,''
  Nucl.\ Phys.\ Proc.\ Suppl.\  {\bf 119} (2003) 71
  [arXiv:hep-ph/0210317].
  %%CITATION = HEP-PH 0210317;%%
See also M.~Knecht and E.~de Rafael,
  %``Patterns of spontaneous chiral symmetry breaking in the large N(c)  limit
  %of QCD-like theories,''
  Phys.\ Lett.\  B {\bf 424} (1998) 335
  [arXiv:hep-ph/9712457].
  %%CITATION = PHLTA,B424,335;%%



\bibitem{Jaffe}
See, e.g., R.~L.~Jaffe,
  %``Ordinary and extraordinary hadrons,''
  AIP Conf.\ Proc.\  {\bf 964} (2007) 1
  [Prog.\ Theor.\ Phys.\ Suppl.\  {\bf 168} (2007) 127]
  [arXiv:hep-ph/0701038] and references therein.
  %%CITATION = PTPSA,168,127;%%


\bibitem{Witten}
E.~Witten,
  %``Baryons In The 1/N Expansion,''
  Nucl.\ Phys.\  B {\bf 160} (1979) 57.
  %%CITATION = NUPHA,B160,57;%%





\bibitem{p6}
 H.~W.~Fearing and S.~Scherer,
  %``Extension of the chiral perturbation theory meson Lagrangian to order
  %p(6),''
  Phys.\ Rev.\  D {\bf 53} (1996) 315
  [arXiv:hep-ph/9408346],
  %%CITATION = PHRVA,D53,315;%%
 J.~Bijnens, G.~Colangelo and G.~Ecker,
  %``The mesonic chiral Lagrangian of order p**6,''
  JHEP {\bf 9902} (1999) 020
  [arXiv:hep-ph/9902437].
  %%CITATION = JHEPA,9902,020;%%

\bibitem{BijnensLis}
M.~Davier, L.~Girlanda, A.~Hocker and J.~Stern,
  %``Finite energy chiral sum rules and tau spectral functions,''
  Phys.\ Rev.\  D {\bf 58} (1998) 096014
  [arXiv:hep-ph/9802447].
  %%CITATION = PHRVA,D58,096014;%%

\bibitem{Pommerenke}
C. Pommerenke, \textit{Pade approximants and convergence in capacity}, J. Math. Anal. Appl. \textbf{41}
(1973) 775. Reviewed in G.A. Baker and P. Graves-Morris, \textit{Pade Approximants}, Encyclopedia of
Mathematics and its Applications, Cambridge Univ. Press 1996, Section 6.5, Theorem 6.5.4, Collorary 1.

\bibitem{Sharpe}
See, e.g.,  S.~R.~Sharpe,
  %``Progress in lattice gauge theory,''
  arXiv:hep-lat/9811006.
  %%CITATION = HEP-LAT/9811006;%%


\bibitem{PDG}
 W.~M.~Yao {\it et al.}  [Particle Data Group],
  %``Review of particle physics,''
  J.\ Phys.\ G {\bf 33} (2006) 1.
  %%CITATION = JPHGB,G33,1;%%

\bibitem{pelaez}
 See, e.g., C.~Hanhart, J.~R.~Pelaez and G.~Rios,
  %``Quark mass dependence of the rho and sigma from dispersion relations and
  %Chiral Perturbation Theory,''
  arXiv:0801.2871 [hep-ph] and references therein.
  %%CITATION = ARXIV:0801.2871;%%
\bibitem{pelaez2}
 J.~R.~Pelaez and G.~Rios,
  %``Scalar mesons from Unitarized Chiral Perturbation Theory: N_c and quark
  %mass dependences,''
  arXiv:0711.4223 [hep-ph] and references therein.
  %%CITATION = ARXIV:0711.4223;%%



\bibitem{Eadie}
W.T. Eadie et al, ``Statistical Methods in Experimental Physics'', North-Holland 1971 and M. Martinez,
private communication.


\bibitem{mp}
 P.~Masjuan and S.~Peris,
  %``A Rational Approach to Resonance Saturation in large-Nc QCD,''
  JHEP {\bf 0705}, 040 (2007)
  [arXiv:0704.1247 [hep-ph]].
  %%CITATION = JHEPA,0705,040;%%

\bibitem{Colangelo}
  G.~Colangelo and S.~Durr,
  %``The pion mass in finite volume,''
  Eur.\ Phys.\ J.\  C {\bf 33} (2004) 543
  [arXiv:hep-lat/0311023].
  %%CITATION = EPHJA,C33,543;%%

  \bibitem{PDG}
  W.~M.~Yao {\it et al.}  [Particle Data Group],
  %``Review of particle physics,''
  J.\ Phys.\ G {\bf 33} (2006) 1.
  %%CITATION = JPHGB,G33,1;%%

\bibitem{Stern}
  M.~Davier, L.~Girlanda, A.~Hocker and J.~Stern,
  %``Finite energy chiral sum rules and tau spectral functions,''
  Phys.\ Rev.\  D {\bf 58} (1998) 096014
  [arXiv:hep-ph/9802447].
  %%CITATION = PHRVA,D58,096014;%%


\bibitem{PGV} C. Diaz-Mendoza, P. Gonzalez-Vera and R. Orive, Appl. Num. Math.
\textbf{53} (2005) 39 and references therein; C. Brezinski and J. Van Iseghem, ``Handbook of Numerical
Analysis'', Vol. III, North Holland 1994.


\bibitem{decays}
 S.~Friot, D.~Greynat and E.~de Rafael,
  %``Chiral condensates, Q(7) and Q(8) matrix elements and large-N(c) QCD,''
  JHEP {\bf 0410} (2004) 043
  [arXiv:hep-ph/0408281] and references therein.
  %%CITATION = JHEPA,0410,043;%%

\bibitem{swiss}
G.~Ecker, J.~Gasser, H.~Leutwyler, A.~Pich and E.~de Rafael,
  %``Chiral Lagrangians for Massive Spin 1 Fields,''
  Phys.\ Lett.\  B {\bf 223} (1989) 425,
  %%CITATION = PHLTA,B223,425;%%
G.~Ecker, J.~Gasser, A.~Pich and E.~de Rafael,
  %``The Role Of Resonances In Chiral Perturbation Theory,''
  Nucl.\ Phys.\  B {\bf 321} (1989) 311.
  %%CITATION = NUPHA,B321,311;%%


\bibitem{run}
 J.~Bijnens, G.~Colangelo and G.~Ecker,
  %``Renormalization of chiral perturbation theory to order p**6,''
  Annals Phys.\  {\bf 280} (2000) 100
  [arXiv:hep-ph/9907333].
  %%CITATION = APNYA,280,100;%%

\bibitem{ABT}
 G.~Amoros, J.~Bijnens and P.~Talavera,
  %``Two-point functions at two loops in three flavour chiral perturbation
  %theory,''
  Nucl.\ Phys.\  B {\bf 568} (2000) 319
  [arXiv:hep-ph/9907264].
  %%CITATION = NUPHA,B568,319;%%

\bibitem{Knecht}
  M.~Knecht and A.~Nyffeler,
  %``Resonance estimates of O(p**6) low-energy constants and QCD  short-distance
  %constraints,''
  Eur.\ Phys.\ J.\  C {\bf 21} (2001) 659
  [arXiv:hep-ph/0106034].
  %%CITATION = EPHJA,C21,659;%%





\bibitem{Portoles}
  V.~Mateu and J.~Portoles,
  %``Form Factors in radiative pion decay,''
  Eur.\ Phys.\ J.\  C {\bf 52} (2007) 325
  [arXiv:0706.1039 [hep-ph]].
  %%CITATION = EPHJA,C52,325;%%
J. Portoles private communication.



\end{thebibliography}
\end{document}